\documentclass[aps,prb,twocolumn,showpacs,superscriptaddress]{revtex4-1}
\usepackage{graphicx}
\usepackage{graphicx}

\begin{document}

\title{Doping driven structural distortion in the bilayer iridate (Sr$_{1-x}$La$_x$)$_3$Ir$_2$O$_7$}

\author{Tom Hogan}
\affiliation{Department of Physics, Boston College, Chestnut Hill, Massachusetts 02467, USA.}
\affiliation{Materials Department, University of California, Santa Barbara, California 93106, USA.}

\author{Xiaoping Wang}
\affiliation{Chemical and Engineering Materials Division, Oak Ridge National Laboratory, Oak Ridge, Tennessee 37831, USA}

\author{H. Chu}
 \affiliation{Department of Physics, California Institute of Technology, Pasadena, California 91125, USA}
 \affiliation{Institute for Quantum Information and Matter, California Institute of Technology, Pasadena, California 91125, USA}

\author{David Hsieh}
 \affiliation{Department of Physics, California Institute of Technology, Pasadena, California 91125, USA}
 \affiliation{Institute for Quantum Information and Matter, California Institute of Technology, Pasadena, California 91125, USA}
 
\author{Stephen D. Wilson}
\email{stephendwilson@engineering.ucsb.edu}
\affiliation{Materials Department, University of California, Santa Barbara, California 93106, USA.}

\date{\today}


\begin{abstract}
Neutron single crystal diffraction and rotational anisotropy optical second harmonic generation data are presented resolving the nature of the structural distortion realized in electron-doped (Sr$_{1-x}$La$_x$)$_3$Ir$_2$O$_7$ with $x=0.035$ and $x=0.071$.  Once electrons are introduced into the bilayer spin-orbit assisted Mott insulator Sr$_3$Ir$_2$O$_7$, previous studies have identified the appearance of a low temperature structural distortion and have suggested the presence of a competing electronic instability in the phase diagram of this material.  Our measurements resolve a lowering of the structural symmetry from monoclinic $C2/c$ to monoclinic $P2_1/c$ and the creation of two unique Ir sites within the chemical unit cell as the lattice distorts below a critical temperature $T_S$.  Details regarding the modifications to oxygen octahedral rotations and tilting through the transition are discussed as well as the evolution of the low temperature distorted lattice as a function of carrier substitution.  
\end{abstract}


\maketitle

\section{Introduction}
Sr$_3$Ir$_2$O$_7$ is the bilayer member of the Ruddlesden-Popper series of strontium iridates, and it is comprised of square lattice sheets of corner sharing IrO$_6$ octahedra stacked within bilayers between rock-salt SrO charge reservoir layers.\cite{subramanian1994single}  This material is a unique example of a spin-orbit Mott insulator \cite{kim2008novel} on the verge of a dimensionality-driven metal to insulator transition.\cite{moon2008dimensionality}  As a result, Sr$_3$Ir$_2$O$_7$ manifests a Mott state in the weak limit---one where the combination of strong spin-orbit coupling, strong crystal field, and moderate on-site Coulomb interaction energy scales stabilizes a small gap (E$_G\approx 130$ meV) \cite{okada2013imaging} antiferromagnetic insulating state where $E_G\approx J$.    

A number of studies have explored the carrier induced response of the spin-orbit assisted Mott (SOM) phase in both Sr$_2$IrO$_4$ \cite{chen2015influence,clancy2014dilute,calder2015evolution,kim2016observation,qi2012spin,de2015collapse} and Sr$_3$Ir$_2$O$_7$ \cite{hogan2015first,dhital2014carrier,li2013tuning,he2015spectroscopic} Ruddlesden-Popper iridates in an effort to realize unconventional metallic states such as high temperature superconductivity.\cite{wang2011twisted}  Due to its charge gap being inherently smaller than Sr$_2$IrO$_4$, electron-doping into Sr$_3$Ir$_2$O$_7$ via La-substitution provides an important example of where the SOM state can be completely quenched and a global metallic state (\textit{i.e.} no nanoscale phase separation) can be realized before the solubility limit of La onto the Sr-sites is reached.\cite{hogan2015first}  Specifically, (Sr$_{1-x}$La$_x$)$_3$Ir$_2$O$_7$ generates a first-order transition into a metallic ground state at a doping level of $6\%$ electrons/Ir ($x=0.04$) which is accompanied by the disappearance of long-range antiferromagnetism.\cite{hogan2015first}  The metallic state realized beyond this transition is unconventional with signatures of strong correlation effects, \cite{ahn2016infrared} dimer formation, \cite{hogan2016disordered,sala2015evidence} negative electronic compressibility, \cite{he2015spectroscopic} and other electronic anomalies.\cite{ding2016pressure,lu2017doping} 

One of the anomalous properties inherent to (Sr$_{1-x}$La$_x$)$_3$Ir$_2$O$_7$ is the appearance of a structural distortion below a characteristic temperature $T_S$.\cite{hogan2015first} This lattice distortion seemingly competes with antiferromagnetism and is suggestive of a secondary signature of an otherwise hidden electronic instability. Optical studies have recently suggested the stabilization of a charge density wave-like instability that appears below $T_S$;\cite{chu2017charge} however little remains known regarding the nature of the structural distortion intertwined with this density wave formation.  

One impediment to determining the details of this lattice transition has been the presence of weak Bragg violations of the nominally orthorhombic $Bbcb$ parent structure of Sr$_3$Ir$_2$O$_7$ reported in earlier neutron diffraction measurements.\cite{dhital2012spin} This ambiguity obscured the parent lattice symmetry of this system and complicated the visualization of the parent structure's evolution across $T_S$.  Lifting this impediment, a recent combined single crystal neutron diffraction, rotational anisotropy optical second harmonic generation (RA-SHG), and density functional theory study has now identified the inherent distortion away from $Bbcb$ in Sr$_3$Ir$_2$O$_7$ as a subtle octahedral tilt out of the basal plane within a lower symmetry monoclinic cell (space group $C2/c$).\cite{hogan2016structural}  Knowing the parent symmetry of the undoped system can now be leveraged to more definitively resolve the nature of the doping driven lattice distortion below $T_S$.        

In this paper, we present a single crystal neutron diffraction and RA-SHG study of the lattice distortion realized in (Sr$_{1-x}$La$_x$)$_3$Ir$_2$O$_7$ as the system enters the low temperature structural phase below $T_S$.  Concentrations on both the insulating ($x=0.035$, $T_S\approx200$ K) and metallic ($x=0.071$, $T_S\approx220$ K) sides of the metal-insulator transition are explored and their lattice structures are determined both at $295$ K ($T>T_S$) and at $100$ K ($T<T_S$).  The high temperature lattice structures for both concentrations are consistent with the recently reported parent $C2/c$ space group of Sr$_3$Ir$_2$O$_7$; however below $T_S$ both concentrations reveal an enhanced out-of-plane tilt of IrO$_6$ octahedra and a lowering of the unit cell symmetry into a primitive centered $P2_1/c$ space group.  The result is a low temperature lattice with two distinct Ir-sites capable of supporting a charge-ordered state---a structure consistent with the recent suggestion of a charge density wave in this system.\cite{chu2017charge} 

\begin{figure}[t]
\includegraphics[scale=0.5]{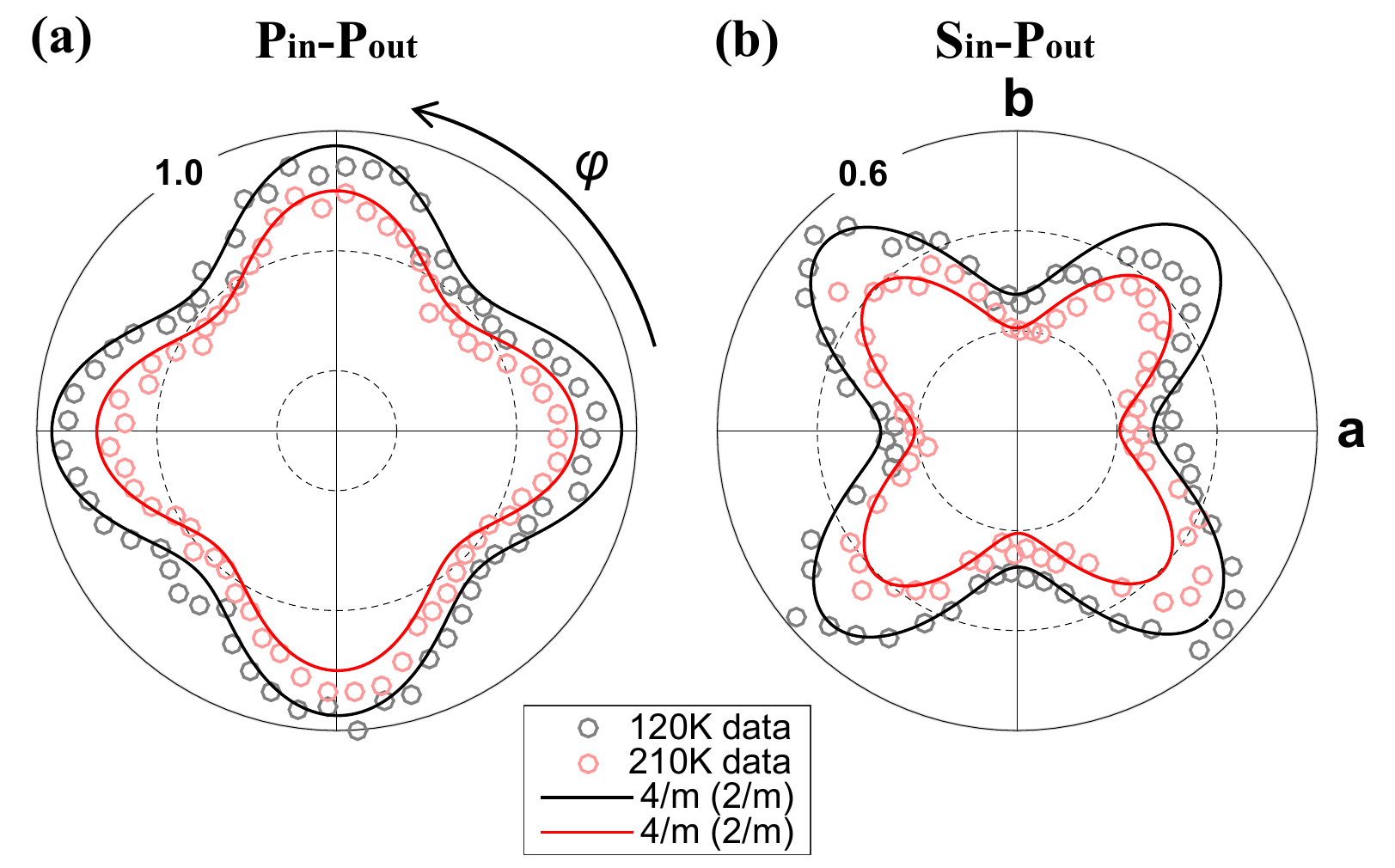}
\caption{Rotational anisotropy second harmonic generation (RA-SHG) patterns of $x=0.033$ (Sr$_{1-x}$La$_{x}$)$_3$Ir$_2$O$_7$ acquired under (a) P$_{in}$-P$_{out}$ and (b) S$_{in}$-P$_{out}$ polarization geometries below (black circles) and above (red circles) the structural phase transition temperature $T_S$. P and S indicate the photon polarization parallel and perpendicular to the scattering plane respectively and $in$ and $out$ designate the incident and scattered light respectively.  All intensities are normalized to the maximum value of the P$_{in}$-P$_{out}$ pattern at $T = 120$ K. Black and red lines are best fits to the low and high temperature data respectively using an expression for bulk electric quadrupole SHG radiation from a centrosymmetric $4/m$ or $2/m$ crystallographic point group. The P$_{in}$-S$_{out}$ and S$_{in}$-S$_{out}$ data\cite{SupplementalMaterials} yield the same conclusions.}
\end{figure}

\section{Experimental Details}
The methods for growing the (Sr$_{1-x}$La$_{x}$)$_3$Ir$_2$O$_7$ single crystals used in this study are reported elsewhere.\cite{hogan2015first, SupplementalMaterials} Neutron scattering datasets were collected at $295$ K and $100$ K for single crystals with La concentrations of $x = 0.035$ (mass 4.9 mg, dimensions 1.5 mm $\times$ 1.44 mm $\times$ 0.48 mm) and $x = 0.071$ (mass 1.3 mg, dimensions 1.34 mm $\times$ 1.10 mm $\times$ 0.24 mm) using the time-of-flight Laue diffractometer TOPAZ at Oak Ridge National Laboratory.  Optimizing sample orientations to maximize reciprocal space coverage was accomplished using the CRYSTALPLAN software.\cite{zikovsky2011crystalplan}  Raw diffraction data were integrated using a 3D ellipsoidal routine \cite{schultz2014integration} then reduced to account for Lorentz corrections and the time-of-flight spectrum.  Numeric absorption corrections were applied using polyhedral models of the individual crystals, and detector efficiencies were accounted for using the ANVRED3 program.\cite{doi:10.1021/ja00316a031}  The final reduced dataset was refined to models using SHELXL.\cite{Sheldrick:sc5010}  Unless otherwise indicated, reciprocal space vectors are indexed using the $C2/c$ monoclinic space group where $H$ is parallel to the long-axis of the unit cell and $K$ is parallel to the unique axis. 

Charge balance was enforced in all structural solutions, and the site occupancies of O and Ir atoms were fixed to unity in the final structural refinements. The site occupancy for O atoms refined to be fully occupied within error for the x=0.035 sample, but varied for the x=0.071 sample if not constrained. Ir atoms refined to have an occupancy close to unity; however free refinement yielded a several percent (typically $2-3\%$) deviation from full occupation, which may arise as an artifact from the high-absorption of Ir in the neutron structure.  This high absorption introduces an added sensitivity to slight deviations of the real crystal shape from the polyhedral absorption models.

RA-SHG data were acquired using a rotating scattering plane technique \cite{harter2015high} from the (100), long axis, cleaved surface of an $x=0.033$ (Sr$_{1-x}$La$_{x}$)$_3$Ir$_2$O$_7$ single crystal. Incident light was provided by a Ti:sapphire regenerative amplifier (100 fs pulse width, 100 kHz repetition rate, 800 nm center wavelength) and focused onto a 50 $\mu m$ diameter spot on the sample with a fluence of $\approx 1$ mJ/cm$^2$.

\begin{figure}[t]
\includegraphics[scale=0.25]{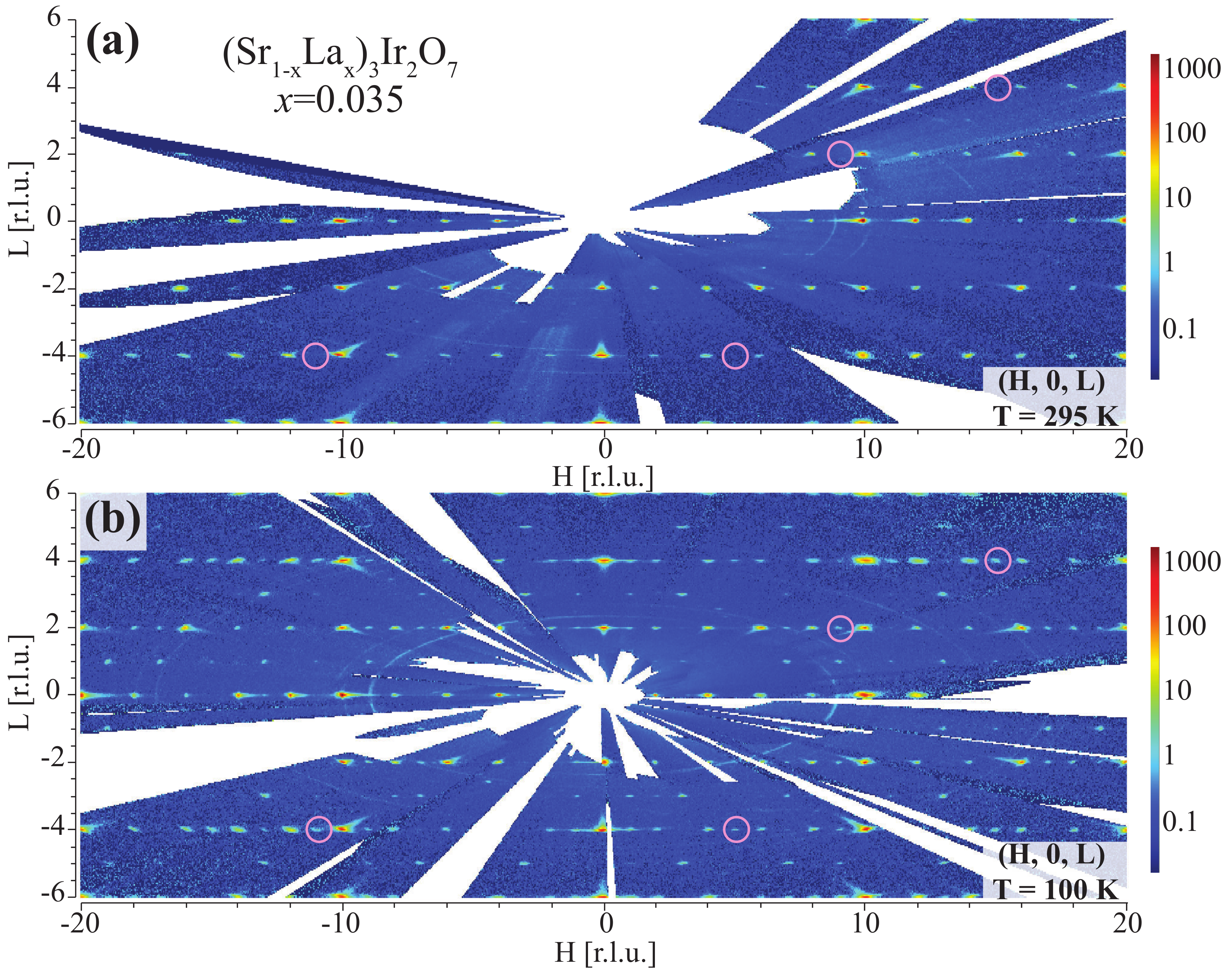}
\caption{Maps of reciprocal space in the [H, 0, L] plane and indexed in using the $C2/c$ space group for the $x=0.035$ sample at both 295 K ($T>T_S$) and at 100 K ($T<T_S$).  Representative violations of the high temperature $C2/c$ space group that appear at 100 K are highlighted by pink circles within the maps.  Peaks of the form $(odd, 0, even)$ violate the twinned scattering conditions for the $C2/c$ cell.}
\end{figure}

\section{RA-SHG Results}

As a first step in resolving the structure of (Sr$_{1-x}$La$_{x}$)$_3$Ir$_2$O$_7$ both below and above $T_S$, RA-SHG measurements were performed to constrain the crystallographic point group symmetry. RA-SHG has been shown to be sensitive to the point group symmetry of a crystal and was applied to help resolve the structure of the parent compound Sr$_3$Ir$_2$O$_7$.\cite{hogan2016structural} Figure 1 shows RA-SHG patterns acquired from the (100) cleaved surface of a $x=0.033$ (Sr$_{1-x}$La$_{x}$)$_3$Ir$_2$O$_7$ single crystal both above and below its structural transition temperature $T_S \approx 200$ K. 

At temperatures above $T_S$, the RA-SHG data in all input and output polarization geometries (parallel P or perpendicular S to the scattering plane) exhibits nearly four-fold rotational symmetry. By using the same analysis applied to undoped Sr$_3$Ir$_2$O$_7$, \cite{hogan2016structural} that is, assuming the dominant contribution to SHG to be of bulk electric-quadrupole origin from a centrosymmetric $4/m$ or $2/m$ point group, we obtain equally good fits to the data. No changes in symmetry are observed below $T_S$ (Fig. 1), only a continuous change in overall magnitude appears upon cooling, and these data demonstrate that the distorted lattice below $T_S$ shares the same crystallographic point group symmetry as the high temperature structure ($T>T_S$). When viewed together with prior diffraction data that excluded space groups with $4/m$ symmetry for the parent lattice structure, these results demonstrate that the point group symmetry remains $2/m$ for (Sr$_{1-x}$La$_{x}$)$_3$Ir$_2$O$_7$ at all temperatures.   

\begin{figure}[t]
\includegraphics[scale=0.35]{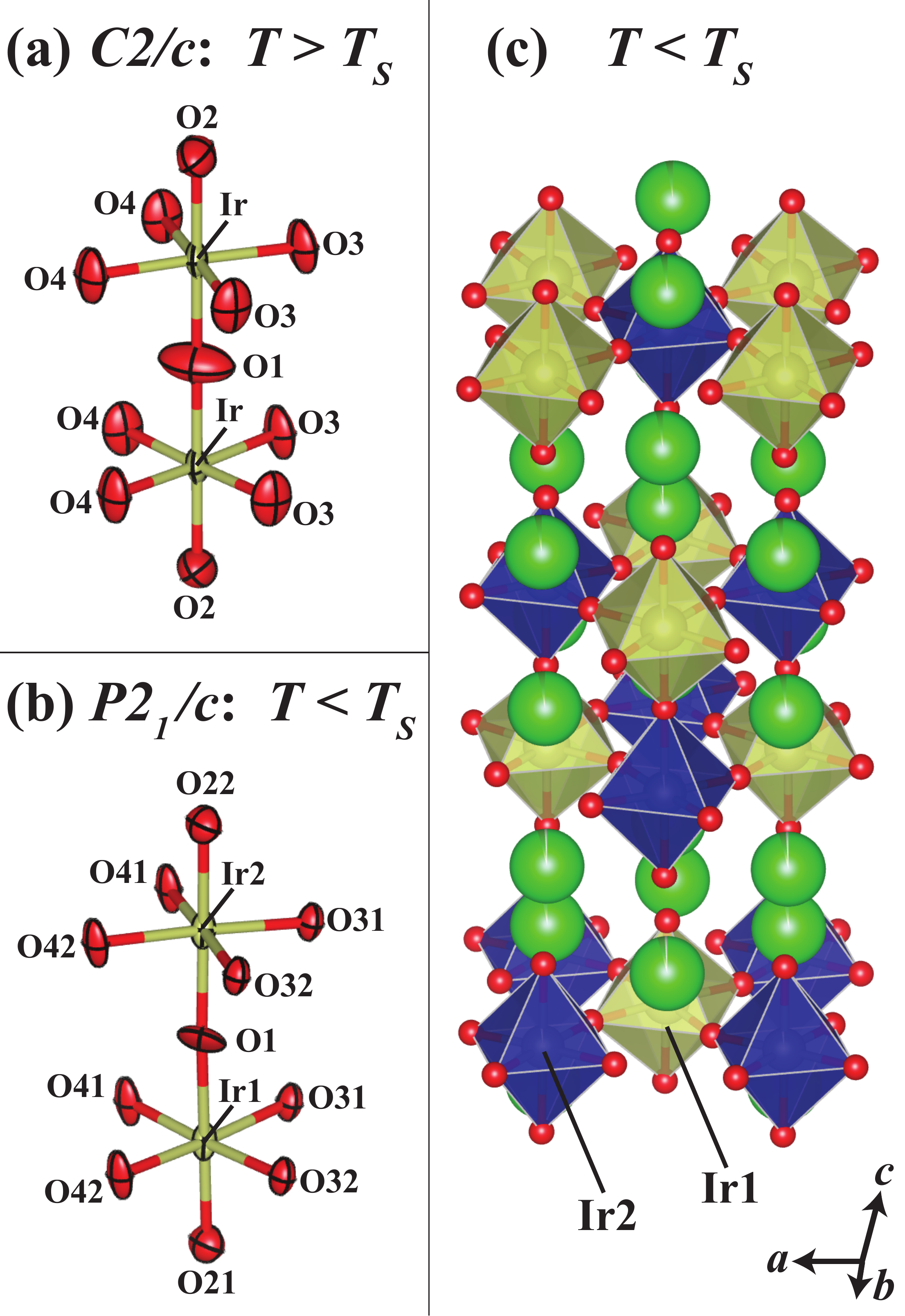}
\caption{Schematic showing the evolution of the crystal structure as the sample is cooled below $T_S$.  Adjacent IrO$_6$ octahedra connected to form a bilayer are shown for (a) $T>T_S$ in the $C2/c$ structure and (b) for $T<T_S$  in the $P2_1/c$ structure. Ir-O1-Ir bonds denote the coupling between the planes of a bilayer. Below $T_S$  basal plane oxygen positions as well as the outer apical O2 oxygen sites split into unique oxygen sites.  Additionally, two chemically distinct Ir-sites appear within a single plane as illustrated by the checkerboard pattern of Ir1 and Ir2 oxygen octahedra illustrated in panel (c).  Green spheres indicate Sr/La positions and red spheres indicate O positions.}
\end{figure}

\section{Neutron Diffraction Results}
Neutron diffraction data were collected on the $x=0.035$ and $x=0.071$ samples at $T=295$ K and $T=100$ K.  The 100 K data reside below $T_S$ for both samples where $T_S \approx 200$ K and $T_S \approx 220$ K for the $x=0.035$ and $x=0.071$ samples respectively. The structures of both $x=0.035$ and $x=0.071$ samples at $T=295$ K refined well within the $C2/c$ space group using the same pseudomerohedral twinning operation as that observed for the $x=0$ system.  This twinning operation involves a two-fold rotation about the basal plane diagonal $(0,1,1)$ direction along the unit cell center, and the resulting twin structure (twins $T_1$ and $T_2$) is effectively just a fault in the phasing of the relative $bc$-plane rotations of IrO$_6$ octahedra between bilayers.  Each twin occupies nearly half of the total scattering volume for both crystals with twinning ratios of $T_1/(T_1+T_2)=0.496$ and $0.492$ refined for $x=0.035$ and $x=0.071$ samples respectively. 

\begin{table}
\caption {Refined lattice parameters and unit cell volumes for (Sr$_{1-x}$La$_x$)$_3$Ir$_2$O$_7$ $x=0$, $x=0.035$, $x=0.071$ at $295$ K and $100$ K.  The space group $C2/c$ (No.\ 15), $Z$ = 4, was used for $295$ K and $x=0$ $100$ K data.  The space group $P2_1/c$ (No.\ 14), $Z$ = 4 was used for $x=0.035$ and $x=0.071$ $100$ K data. $x=0$ data are originally reported elsewhere.\cite{hogan2016structural}}
\begin{ruledtabular}
\begin{tabular}{cccc}
  (Sr$_{1-x}$La$_x$)$_3$Ir$_2$O$_7$ &$x=0$ &$x=0.035$ &$x=0.071$\\
 &($295$ K)&($295$ K)&($295$ K)\\
 \hline
   a (\AA)      &20.935(4)   &20.8877(17)   &20.943(4)   \\
   b (\AA)      &5.5185(13)   &5.5206(5)  &5.5244(16)    \\
   c (\AA)      &5.5099(9)   &5.5287(6)  &5.5138(16)   \\
   $\beta$ ($^{\circ}$)    &90.045(18)   &90.066(8)  &90.05(2)  \\
   V (\AA$^3$)   &636.6(2)   &637.53(10)  &637.9(3)    \\
   
 \hline
(Sr$_{1-x}$La$_x$)$_3$Ir$_2$O$_7$ &$x=0$ &$x=0.035$ &$x=0.071$\\
 &($100$ K)&($100$ K)&($100$ K)\\
 \hline
  a (\AA)      &20.917(3) &5.5133(5)   &5.5109(8)    \\
   b (\AA)      &5.5080(10) &5.5116(5)    &5.5110(8)    \\
   c (\AA)      &5.4995(7) &21.618(2)   &21.611(3)   \\
   $\beta$ ($^{\circ}$)    &90.069(15) &104.821(8)   &104.772(13)  \\
   V (\AA$^3$)   &633.60(17) &635.05(10)   &634.65(16)    \\
\end{tabular}
\end{ruledtabular}
\end{table}

The 295 K lattice parameters and unit cell parameters for the $x=0$, $x=0.035$, and $x=0.071$ samples are summarized in Table I.  Consistent with earlier data, the unit cell volume swells slightly with La-substitution,\cite{hogan2015first} and this expansion is driven primarily by the increase of the basal $bc$-plane lattice parameters.  Atomic positions and anisotropic thermal displacement parameters for $x=0.035$ and $x=0.071$ at 295 K are summarized in Tables II and III respectively.  La concentrations for both samples refine within error to the nominal values determined via earlier energy dispersive x-ray spectroscopy measurements.  

At 295 K, the local ligand field at the Ir sites remains very nearly cubic with small distortion parameters of $\Delta_{d} = \frac{1}{6} \sum_{n=1,6}\big[ (d_n-d)/d \big]^{2}=8.77 \times 10^{-5}$ for $x=0.035$ and $\Delta_{d} = 12.12 \times 10^{-5}$ for $x=0.071$, where $d$ is the average of the octahedral bond lengths.  Basal plane ($bc$-plane) rotation angles of the IrO$_6$ octahedra for both La-doped samples are similar to the parent system with canting angles $\frac{180-\Theta}{2}=11.95^{\circ}$ for $x=0.035$ and $\Theta=11.85^{\circ}$ for $x=0.071$.  The $\Theta$ angles here represent averages of the two distinct Ir-O-Ir bond angles in the $bc$-plane due to the unique O(3) and O(4) sites of the $C2/c$ structure. Individually, these values were $\Theta_{Ir-O(4)-Ir}=156.1^{\circ}$ and $\Theta_{Ir-O(3)-Ir}=156.1^{\circ}$ for $x=0.035$ and $\Theta_{Ir-O(4)-Ir}=156.2^{\circ}$ and $\Theta_{Ir-O(3)-Ir}=156.4^{\circ}$ for $x=0.071$.  Out-of-plane tilting angles of octahedra between bilayers remain small with values of $\frac{180-\Gamma}{2}= 0.1^{\circ}$ for $x=0.035$ and $0.15^{\circ}$ for $x=0.071$ where $\Gamma$ is the Ir-O(1)-Ir interplane bond angle.  This verifies that, in the high temperature phase, the structure is largely unperturbed by the substitution of La into the lattice.

\begin{table}
\caption{Results of refinement of $295$ K, $x=0.035$ neutron diffraction data to the $C2/c$ model.  Wyckoff site labels, relative atomic coordinates and anisotropic displacement factor matrices $U_{ij}$ are included.  $U_{ij}$ values in the table are multiplied by $10^3$.  $R1= 0.057$}
\begin{ruledtabular}
\begin{tabular}{lccccc}
 Atom &Site &x &y &z &U$_{11}$ \\
 \hline

		Ir &8f   &0.59774(4)  &0.7499(3) &0.74997(13) &9.7(4)\\
		Sr(1) &4e   &0.5000      &0.2498(6) &0.7500 &11.3(7)		\\
		La(1) &4e   &0.5000      &0.2498(5) &0.7500 &11.3(7)		\\
		Sr(2) &8f   &0.68706(6)  &0.7497(5) &0.2497(3) &10.2(6)	\\
		La(2) &8f   &0.68706(6)  &0.7497(5) &0.2497(3) &10.2(6)	\\
		O(1) &4e   &0.5000      &0.7506(9) &0.7500  &10.0(10)	\\
		O(2) &8f   &0.69492(8)  &0.7492(6) &0.7498(3) &10.1(6) 	\\
		O(3) &8f   &0.09607(11) &0.4470(4) &0.4474(4) &23.4(10)	\\
		O(4) &8f   &0.09592(11) &0.9473(4) &0.5525(4) &22.2(9)	\\

 \hline
 Atom &U$_{22}$ &U$_{33}$ &U$_{23}$ &U$_{13}$ &U$_{12}$\\
 \hline

		Ir      		&8.5(18) &2.9(15) &0.0(3) &-0.6(3)  &-0.2(3)	\\
		Sr(1)   	&5(4) 	&13(4) 	&0.00 &-0.7(6)  &0.00       \\
		La(1)  	&5(4) 	&13(4) 	&0.00 &-0.7(6)  &0.00       \\
		Sr(2)   	&10(3) 	&11(2) 	&-0.6(6) &-0.5(4)  &-0.1(6)  \\
		La(2)   	&10(3) 	&11(2) 	&-0.6(6) &-0.5(4)  &-0.1(6)  \\
		O(1)   	&17(5) 	&41(5) 	&0.00    &-1.6(11) &0.00       \\
		O(2)    	&18(2) 	&9.7(16) &-0.5(7) &-0.4(5)  &-0.7(8)  \\
		O(3)    	&7.1(9) &8.8(8) 	&2.6(8) 	&0.3(8) 	 &0.3(9)   \\
		O(4)    	&6.8(8) 	&9.1(8) 	&-3.1(8) &-1.4(8)  &-0.1(9)  \\
	
\end{tabular}
\end{ruledtabular}
\end{table}

\begin{table}
\caption{Results of refinement of $295$ K, $x=0.071$ neutron diffraction data to the $C2/c$ model.  Wyckoff site labels, relative atomic coordinates and anisotropic displacement factor matrices $U_{ij}$ are included.   $U_{ij}$ values in the table are multiplied by $10^3$. $R1= 0.046$}
\begin{ruledtabular}
\begin{tabular}{lccccc}
 Atom &Site &x &y &z &U$_{11}$  \\
 \hline

		Ir     &8f	&0.59773(4) 	&0.7504(8) 	&0.74983(16) &8.6(4)\\
		Sr(1)  &4e &0.5000 		&0.2507(2) 	&0.7500 &10.1(7) 		\\
		La(1)  &4e &0.5000 		&0.2507(2) 	&0.7500 &10.1(7)		\\
		Sr(2)  &8f 	&0.68692(6) 	&0.7488(15) 	&0.2502(3) &9.5(5)\\
		La(2)  &8f 	&0.68695(7) 	&0.7510(19) 	&0.2502(3) &9.5(5)\\
		O(1)   &4e	  &0.5000 		&0.749(2) 		&0.7500 &7.6(8)		\\
		O(2)   &8f	 &0.69475(7) 	&0.7499(12) 	&0.7501(3) &8.7(6)\\
		O(3)   &8f	 &0.09600(13) 	&0.4481(6) 	&0.4479(6) &21.4(10)\\
		O(4)   &8f	 &0.09595(13) 	&0.9479(6) 	&0.5526(6) &22.8(10)\\

 \hline
 Atom &U$_{22}$ &U$_{33}$ &U$_{23}$ &U$_{13}$ &U$_{12}$\\
 \hline

		Ir    	 &9.5(18) 	&0.0026(14) 	&0.4(5) 	&-1.4(3) 	&0.4(9)\\
		Sr(1) &6(4) 	&12(4) 	&0.00 		&-2.2(6) 	&0.00      	\\
		La(1) &6(4) 	&12(4) 	&0.00 		&2.2(6)	&0.00      	\\
		Sr(2) &19(2) 	&2.8(18) 	&-1.4(10) 	&1.0(4)	&1.3(19)  	\\
		La(2) &19(2) 	&2.8(18)	 &-1.4(10) 	&1.0(4) 	&1.3(19)  	\\
		O(1)  &23(7) 	&33(8) 	&0.00 		&0.0(10) 	&0.00      	\\
		O(2)  &9(4) 	&18(5) 	&0.2(15) 	&-1.7(6) 	&3(2)  	\\
		O(3)  &9.6(15) 	&8.0(13) 	&2.8(13) 	&-1.7(16) 	&1.6(19)  	\\
		O(4)  &7.1(15)	&5.8(12) &-1.0(13) 	&-0.5(16) 	&0.3(19)   	\\
	
\end{tabular}
\end{ruledtabular}
\end{table}

Upon cooling to 100 K, a number of new Bragg peaks appear in the single crystal diffraction patterns of both the $x=0.035$ and $x=0.071$ crystals.   These new peaks appear at positions that violate the general reflection conditions of space group $C2/c$ that for $[H,0,L]$: $H,L = 2n$ and for $[H,K,0]$: $H+K = 2n$.  Bragg peaks collected in the $[H,0,L]$ plane for the $x=0.035$ sample are plotted in Fig. 2 at both $300$ K and $100$ K.  For illustration, four peaks are highlighted at the (-11, 0, -4), (5, 0, -4), (9, 0, 2), and (15, 0, 4) positions---all of which appear only in the 100 K pattern.  Even after accounting for twin domains by transposing the $K, L$ indices via the twin operator (1 0 0, 0 0 -1, 0 1 0) where the triplets indicate matrix rows, these reflections violate the $C2/c$ space group conditions.  This demonstrates a further lowering of the lattice symmetry through $T_S$ rather than a simple enhancement of the existing, subtle distortion inherent to the parent Sr$_3$Ir$_2$O$_7$'s $C2/c$ structure.  

In determining the likely space groups for the structure below $T_S$, the fact that the transition is second order reported from earlier neutron order parameter measurements \cite{hogan2015first} mandates that the lower symmetry group should be a subgroup of the parent $C2/c$ phase (No.\ 15).  In looking to lower symmetry however, Bragg peaks appearing below $T_S$ violate the general reflection conditions of even the lowest-symmetry $C$-centered space group $C2$ (No.\ 5), which requires for $[H,0,L]$: $H = 2n$.  This demonstrates that the lattice is no longer base-centered at low temperature.   Space groups which preserve the point group symmetry ($2/m$) but instead have primitive cells are Nos.\ 14, 13, 11, and 10.  Within this list, Nos. 10 and 11 have no group-subgroup relationship with No.\ 15, leaving Nos.\ 14 and 13 as likely candidates for describing the lattice of the distorted low temperature phase. Both space group Nos. 14 and 13 were attempted, and superior fits were obtained within the higher symmetry group $P2_1/c$.  The loss of a two-fold rotation axis allows for an oxygen octahedral distortion that creates three additional unique oxygen sites as well as an additional chemically distinct Ir position within the IrO$_6$ planes.  

It is worth briefly describing this new cell setting, illustrated in Fig. 3.  In $P2_1/c$, the long axis of the cell is the $c$-axis; however with a cell $\beta \approx 105^{\circ}$ which corresponds to shifting over one basal plane lattice constant as the unit cell is traversed along the long-axis in a $\beta \approx 90^{\circ}$ setting.  The indices of the $C2/c$ cell are related to the new primitive $P2_1/c$ cell by the transform (0 0 -1, 0 1 0, 1 0 -1).  For refinements within this $P2_1/c$ cell, the program ROTAX \cite{cooper2002derivation} was used to identify the most likely twin operator as a four-fold rotation about the $[1,0,1]$ direction of the primitive cell.  This direct-space lattice direction corresponds precisely to the long-axis of the base-centered unit-cell in $C2/c$ and the four-fold rotation about such an axis in the tetragonal metric is a symmetry operator of the $4/mmm$ point group, classifying this also as twinning by pseudomerohedry.  

The new primitive monoclinic unit cell combined with the twinning operation (0 1 0, -1 0 0, 1 -1 1) accounts for all of the disallowed reflections observed below $T_S$, where there are now no restrictions on reflection conditions.  The refined weights of the contributions of individual twin-domains are each very near the expected value of 0.25 for a four-fold symmetric twin operator with $T_1=0.241$,   $T_2=0.248$,  $T_3= 0.263$, $T_4=0.248$ for the $x=0.035$ sample and $T_1=0.216$,   $T_2=0.257$,   $T_3=0.279$, $T_4=0.248$ for the $x=0.071$ sample.  A summary of the 100 K unit cell parameters for both samples is provided in Table I.

\begin{figure}[t]
\includegraphics[scale=0.2]{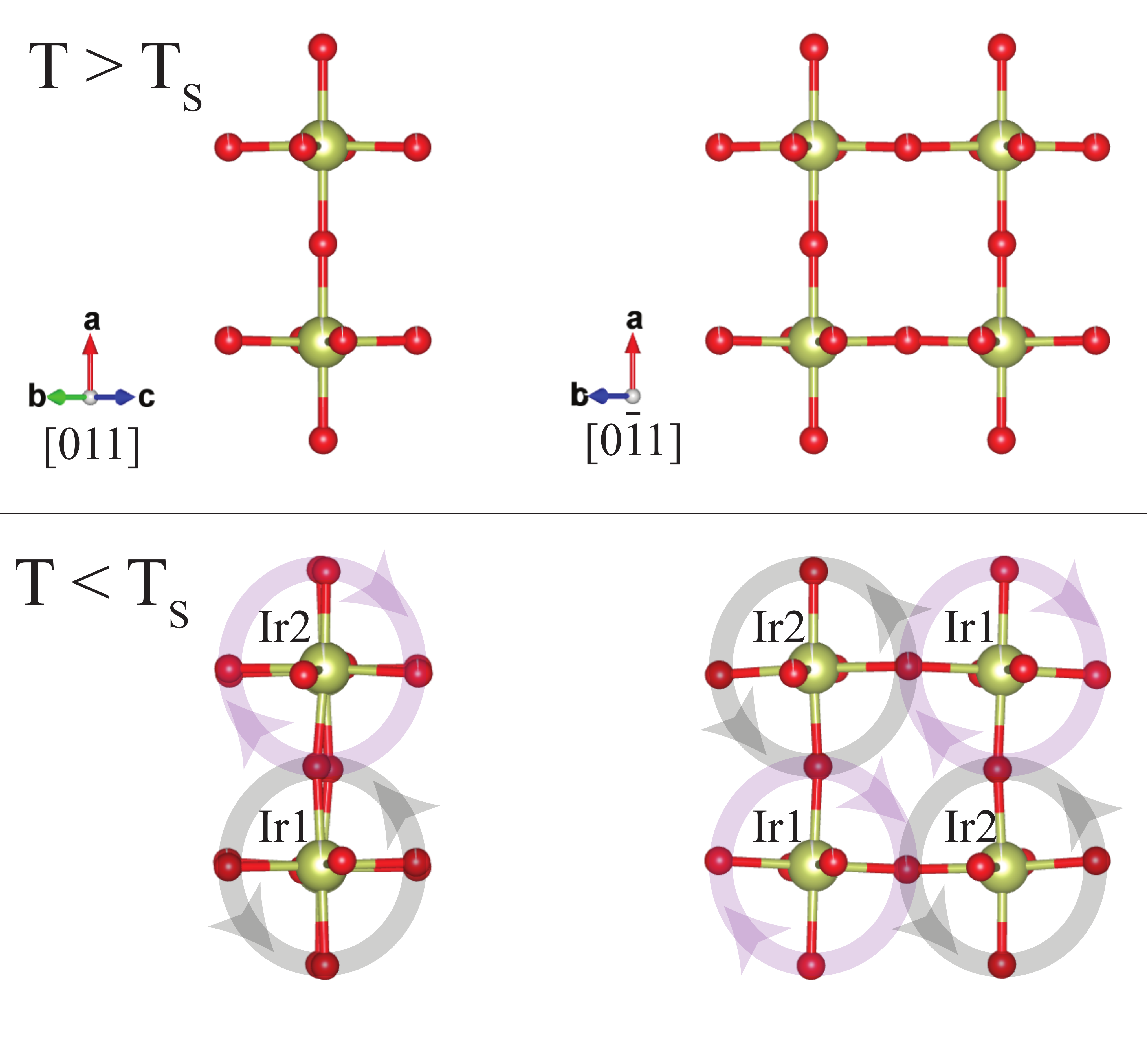}
\caption{Rotational pattern of IrO$_6$ octahedral tilting within a bilayer of the $x=0.035$ crystal activated below $T_S$.  The sense of correlated titling is illustrated looking down the $[0, 1, 1]$ and $[0, \bar{1}, 1]$ axes of the $C2/c$ unit cell where nearest neighbor IrO$_6$ octahedra tilt with an opposite sense relative to one another.}
\end{figure}

A good structural refinement with anisotropic displacement parameters was obtained at 100 K using the twinned $P2_1/c$ cell for the $x=0.035$ sample. Atomic positions and the resulting displacement parameters are listed in Table IV. In the lower symmetry $P2_1/c$ cell, the majority of atomic sites split into two distinct positions, and a stable refinement with anisotropic thermal parameters required the constraint that the displacement parameters of sites split through the transition from $C2/c$ to $P2_1/c$ be fixed identical to one another.  We again note here that this lower symmetry cell has two distinct Ir-sites, labeled Ir(1) and Ir(2) (illustrated in Fig. 3).  The local oxygen octahedral environment of each Ir-site becomes slightly more distorted relative to the high temperature 295 K structure with distortion parameters $\Delta_{d}=18.45 \times 10^{-5}$ and  $\Delta_{d}=13.34 \times 10^{-5}$ for the Ir(1) and Ir(2) sites respectively.  Nearest neighbor Ir-Ir distances within the basal plane and interbilayer Ir-Ir distances remain unchanged within error.

The average  basal plane Ir-O-Ir bond angle decreases slightly to $155.36^{\circ}$ at 100 K in the $x=0.035$ sample, resulting in an average increase in the octahedral canting angle to $12.3^{\circ}$.  There is however greater variance in the basal plane Ir-O-Ir bond angles, where $\Theta_{Ir(1)-O(42)-Ir(2)}=157.4^{\circ}$, $\Theta_{Ir(1)-O(32)-Ir(2)}=153.7^{\circ}$, $\Theta_{Ir(1)-O(31)-Ir(2)}=154.7^{\circ}$, and $\Theta_{Ir(1)-O(41)-Ir(2)}=155.6^{\circ}$.   Above $T_S$, the bond angles Ir-O(3)-Ir and Ir-O(4)-Ir are identical to one another.  

\begin{table}
\caption{Results of refinement of $100$ K, $x=0.035$ neutron diffraction data to the $P2_1/c$ model.  Wyckoff site labels, relative atomic coordinates and anisotropic displacement factor matrices $U_{ij}$ are included.   $U_{ij}$ values in the table are multiplied by $10^3$. $R1= 0.053$}
\begin{ruledtabular}
\begin{tabular}{lccccc}
 Atom &Site &x &y &z &U$_{11}$  \\
 \hline

      Ir(1) 	&4e   &0.8433(4) &0.2468(4) &0.59748(13) &5.1(7)\\
      Ir(2) 	&4e  &0.6571(4) &0.2534(4) &0.40207(13) &5.1(7)\\
      Sr(1) 	&4e  &0.2542(5) &0.2459(13) &0.5004(3)   &4.0(12)\\
      Sr(21) 	&4e  &0.9399(6) &0.7390(8) &0.6863(2) &7.7(12)\\	
      Sr(22) &4e  &0.5593(6) &0.7670(8) &0.3122(2)  &7.4(11)\\	
      La(1) 	&4e  &0.2542(5) &0.2459(13) &0.5004(3)   &4.0(12)\\
      La(21) 	&4e  &0.9399(6) &0.7390(8) &0.6863(2)   &7.7(12)\\	
      La(22) &4e  &0.5593(6) &0.7670(8) &0.3122(2)   &7.4(11)\\	
      O(1) 	&4e  &0.7248(9) &0.2498(17) &0.4994(3)   &15.6(15)\\
      O(21) 	&4e  &0.9555(7) &0.2449(12) &0.6947(3)   &9.0(9)\\
      O(22) 	&4e  &0.5671(7) &0.2555(12) &0.3050(2)   &9.0(9)\\
      O(31) 	&4e  &0.1416(6) &0.0482(7) &0.59123(17) &5.9(6)\\
      O(32) 	&4e  &0.3459(6) &0.0593(7) &0.4021(2)   &5.9(6)\\
      O(41) 	&4e  &0.0382(7) &0.5516(7) &0.5942(2)   &5.3(6)\\
      O(42) 	&4e  &0.4510(6) &0.5550(7) &0.39938(19)   &5.3(6)\\
 \hline
 Atom &U$_{22}$ &U$_{33}$ &U$_{23}$ &U$_{13}$ &U$_{12}$\\
 \hline

		Ir(1) 		    &1.9(5) &7.61(14) &-0.13(10) &1.0(5) &-0.7(3) \\
		Ir(2) 		    &1.9(5) &7.61(14) &-0.13(10) &1.0(5) &-0.7(3)  \\
		Sr(1) 	    &4.0(12) &8.6(3) &0.7(3) &1.2(10) &-0.6(11)   \\
		Sr(21) 	    &4.1(11) &9.6(12) &0.6(9) &5.4(8) &-1.5(7)   \\
		Sr(22) 	    &6.3(12) &6.3(10) &-0.3(9) &4.5(7) &-1.7(7)   \\
		La(1) 	    &4.0(12) &8.6(3) &0.7(3) &1.2(10) &-0.6(11)   \\
		La(21) 	    &4.1(11) &9.6(12) &0.6(9) &5.4(8) &-1.5(7)   \\
		La(22) 	    &6.3(12) &6.3(10) &-0.3(9) &4.5(7) &-1.7(7)   \\
		O(1) 		    &17.2(13) &9.4(5) &-0.1(5) &7.7(15) &2(2)  \\
		O(21) 	    &9.0(8) &8.2(2) &0.2(2) &1.8(8) &1.3(11)   \\
		O(22) 	    &9.0(8) &8.2(2) &0.2(2) &1.8(8) &1.3(11)   \\
		O(31) 	    &4.3(8) &11.6(8) &1.5(8) &1.8(6) &0.4(5)    \\
		O(32) 	    &4.3(8) &11.6(8) &1.5(8) &1.8(6) &0.4(5)    \\
		O(41)           &6.6(8) &16.0(10) &-1.5(9) &4.8(6) &0.7(6)    \\
		O(42) 	    &6.6(8) &16.0(10) &-1.5(9) &4.8(6) &0.7(6)     \\	
\end{tabular}
\end{ruledtabular}
\end{table}

The most prominent distortion to the lattice structure below $T_S$ is the activation of interlayer apical tilting.  The Ir(1)-O(1)-Ir(2) bond angle is appreciably distorted relative to the high temperature structure and reduces to $172.3^{\circ}$, resulting in a $3.9^{\circ}$ octahedral tilting angle.  This activated tilting of the oxygen octahedra about both the basal plane $(0,1,1)$ and $(0,\bar{1},1)$ vectors in the $C2/c$ cell represents the primary distinction between the structures below and above $T_S$.  The relative senses of tilts of the oxygen octahedra along this apical bond are illustrated in Fig. 4. 

For the $x=0.071$ sample at 100 K, a similar approach to refining the low temperature unit cell in $P2_1/c$ was attempted;\cite{SupplementalMaterials} however anisotropic displacement parameters for a number of sites failed to converge to positive definite values.  The $x=0.071$ crystal was found to possess a small second grain closely aligned with the primary domain which likely triggered this failure.  Nevertheless when both domains were integrated, a refinement employing isotropic displacement parameters was successful in $P2_1/c$ with atomic site positions and thermal factors tabulated in Table V.  The distorted structure for this metallic sample is qualitatively similar to that observed in the $x=0.035$ sample.   The Ir(1)-O(1)-Ir(2) out of plane bond angle was refined to be $169.9^{\circ}$ ($5.1^{\circ}$ tilting angle), a value slightly larger than that determined via the anisotropic refinement of the $x=0.035$ sample.  This apparent enhancement of the octahedral tilt relative to the $x=0.035$ sample is likely due to the isotropic refinement of the lattice, and an isotropic refinement of the $x=0.035$ 100 K data yields a similarly enhanced tilting angle of $4.7^{\circ}$.  
 
\begin{table}
\caption{Results of refinement of 100 K, $x=0.071$ neutron diffraction data to the $P2_1/c$ model.  Wyckoff site labels, relative atomic coordinates and isotropic displacement factors $U_{iso}$ are included.  $U_{ij}$ values in the table are multiplied by $10^3$. $R1= 0.097$}
\begin{ruledtabular}
\begin{tabular}{lccccc}
 Atom &Site &x &y &z &$U_{iso}$  \\
 \hline

   Ir(1) 	&4e 	&0.8452(11) &0.2467(16) &0.59526(14) &8.3(2) 	\\
   Ir(2) &4e 	&0.6499(11) &0.2547(17) &0.39994(14) &8.3(2) 	\\
   Sr(1) 	&4e 	&0.2561(14) &0.244(3) &0.5028(3) &8.2(6)	\\
   Sr(21) 	&4e 	&0.9343(11) &0.7333(17) &0.6845(2) &5.8(9)	\\
   Sr(22) &4e 	&0.5620(13) &0.764(2) &0.3099(3) &9.7(11)	\\
   La(1) 	&4e 	&0.2561(14) &0.244(3) &0.5028(3) &8.2(6)	\\
   La(21) 	&4e &0.9343(11) &0.7333(17) &0.6845(2) &5.8(9)	\\
   La(22) &4e 	&0.5620(13) &0.764(2) &0.3099(3) &9.7(11)	\\
   O(1) 	&4e		&0.7178(16) &0.241(3) &0.4992(5) &11.7(8)	\\
   O(21) 	&4e		&0.9707(15) &0.2445(16) &0.6971(3) &9.5(5)	\\
   O(22) 	&4e		&0.5484(15) &0.2545(16) &0.3069(3) &9.5(5)	\\
   O(31) 	&4e		&0.1516(19) &0.0507(18) &0.5968(3) &8.5(3)	\\
   O(32) 	&4e		&0.3476(17) &0.0587(16) &0.4049(3) &8.5(3) 	\\
   O(41) 	&4e		&0.0364(16) &0.5501(17) &0.5883(3) &8.5(3) 	\\
   O(42) 	&4e		&0.4481(15) &0.5554(16) &0.3963(3) &8.5(3) 	\\

\end{tabular}
\end{ruledtabular}
\end{table}

The average oxygen octahedral distortion parameters about Ir-sites increase below $T_S$ for this $x=0.071$ crystal with $\Delta_{d} = 60.42 \times 10^{-5}$ for the octahedra about the Ir(1) sites and $\Delta_{d} = 42.21 \times 10^{-5}$ for octahedra about the Ir(2) sites.  A similar increase in $\Delta_{d}$ below $T_S$ is not observed in either an anisotropic or isotropic refinement of the 100 K $x=0.035$ data, suggesting that this effect is not purely a consequence of the isotropic treatment of the $x=0.071$ data.  Similar to the $x=0.035$ sample, the average basal plane IrO$_6$ octahedral rotation angle increases slightly relative to the high temperature 295 K value (from $11.85^{\circ}$ to $12.61^{\circ}$).       

\section{Discussion}
The manifestation of a global lattice distortion under very light La-doping levels combined with its competition with the antiferromagnetic insulating state within the phase diagram of (Sr$_{1-x}$La$_x$)$_3$Ir$_2$O$_7$ \cite{hogan2015first} is suggestive of its being a secondary consequence of a primary electronic order parameter.  Within this scenario, the distortion observed below $T_S$ in La-substituted Sr$_3$Ir$_2$O$_7$ is consistent with the recent suggestion of charge density wave-like order near the same temperature.\cite{chu2017charge}  For such a scenario to occur, the structure should accommodate any additional symmetry breaking driven via charge order on the Ir-sites via the creation of chemically distinct Ir-sites compatible with hosting charge texture within the lattice.  

The distortion from $C2/c$ to $P2_1/c$ observed below $T_S$ allows for potential charge disproportionation on the Ir sites through the creation of distinct Ir(1) and Ir(2) sites arranged in a checkerboard pattern within the IrO$_2$ planes of a given bilayer.  This is reminiscent of previous observations of structural accommodation of charge order within perovskite nickelates \cite{hwang2013structural} and supports a picture of a competing charge instability in this electron-doped SOM insulator.  We note here that the formation of two Ir-sites in electron-doped Sr$_3$Ir$_2$O$_7$ is distinct from the two Ir-sites inherent to the Sr$_2$IrO$_4$ lattice.\cite{torchinsky2015structural}  In single layer Sr$_2$IrO$_4$, the two unique Ir sites are an intrinsic property of the parent crystal structure,\cite{ye2013magnetic,dhital2013neutron,torchinsky2015structural} and there is no structural distortion evident upon electron doping.\cite{chen2015influence}   

The appreciable tilting of the IrO$_6$ octahedra out of the basal plane is the primary signature of the lattice distortion below $T_S$ and, within the uncertainty of our current refinement, the magnitudes of these activated tilts are equivalent for the $x=0.071$ and $x=0.035$ samples.  While this comparison is somewhat constrained by the inability to perform a refinement with anisotropic displacement parameters for the  $x=0.071$ sample, the observation is consistent with previous data demonstrating that the magnitude of the competing lattice distortion saturates at the insulator to metal transition near $x=0.035$ in the phase diagram of (Sr$_{1-x}$La$_x$)$_3$Ir$_2$O$_7$.\cite{hogan2015first}  

The enhanced distortion of the local oxygen octahedra about the Ir sites in the $x=0.071$ crystal relative to the $x=0.035$ crystal marks one distinction between the two low temperature structures.   An isotropic refinement of the $x=0.071$ sample shows an alternating enhancement/contraction of the unshared apical Ir-O(2) bond lengths as the system is cooled through $T_S$.  In the low temperature phase, the Ir(1)-O(21) bond length expands by $\approx 0.1$ $\AA$ while the Ir(2)-O(22) bond length contracts by $\approx 0.1$ $\AA$.  While this is naively consistent with a partial charge disproportionation on the Ir(1) and Ir(2) sites that builds with continued electron doping, future measurements of a highly doped (Sr$_{1-x}$La$_x$)$_3$Ir$_2$O$_7$ crystal (without the complication of a small second grain) that allow for a full refinement of anisotropic displacement parameters are required to more carefully explore this possibility.  

\section{Conclusions}
In summary, we have performed single crystal neutron diffraction measurements resolving the nature of the lattice distortion that manifests across the structural transition $T_S$ in the electron-doped SOM insulator (Sr$_{1-x}$La$_x$)$_3$Ir$_2$O$_7$.  The high temperature ($T>T_S$) cell retains its $C2/c$ symmetry through the solubility limit of La ($x\approx0.07$) into the lattice.  The low temperature ($T<T_S$) unit cells for both the $x=0.035$ concentration near the metal to insulator phase boundary and for the $x=0.071$ concentration deep in the metallic regime adopt a lower symmetry $P2_1/c$ space group where an appreciable out-of-plane IrO$_6$ octahedral tilt is activated.  Two unique Ir sites appear in this lower symmetry cell, consistent with a lattice symmetry required for the onset of charge density ordering below $T_S$. Our observations support the notion of the lattice distortion manifesting as a secondary consequence of a primary electronic instability in the phase diagram of (Sr$_{1-x}$La$_x$)$_3$Ir$_2$O$_7$.     
\newline
\acknowledgments{
This work was supported by NSF Award No. DMR-1505549 (S.D.W., T.H.), as well as by the Institute for Quantum Information and Matter, an NSF Physics Frontiers Center (PHY-1125565) with support of the Gordon and Betty Moore Foundation through Grant GBMF1250 (D.H., H.C.). Work performed at the ORNL Spallation Neutron Source TOPAZ single-crystal diffractometer was supported by the Scientific User Facilities Division, Office of Basic Energy Sciences, US Department of Energy, under Contract No. DE-AC05-00OR22725 with UT-Battelle, LLC.  VESTA software was used for crystal structure visualization.\cite{Momma2011}}

\bibliography{LaSr327}

\begin{thebibliography}{35}
\expandafter\ifx\csname natexlab\endcsname\relax\def\natexlab#1{#1}\fi
\expandafter\ifx\csname bibnamefont\endcsname\relax
  \def\bibnamefont#1{#1}\fi
\expandafter\ifx\csname bibfnamefont\endcsname\relax
  \def\bibfnamefont#1{#1}\fi
\expandafter\ifx\csname citenamefont\endcsname\relax
  \def\citenamefont#1{#1}\fi
\expandafter\ifx\csname url\endcsname\relax
  \def\url#1{\texttt{#1}}\fi
\expandafter\ifx\csname urlprefix\endcsname\relax\def\urlprefix{URL }\fi
\providecommand{\bibinfo}[2]{#2}
\providecommand{\eprint}[2][]{\url{#2}}

\bibitem[{\citenamefont{Subramanian et~al.}(1994)\citenamefont{Subramanian,
  Crawford, and Harlow}}]{subramanian1994single}
\bibinfo{author}{\bibfnamefont{M.}~\bibnamefont{Subramanian}},
  \bibinfo{author}{\bibfnamefont{M.}~\bibnamefont{Crawford}}, \bibnamefont{and}
  \bibinfo{author}{\bibfnamefont{R.}~\bibnamefont{Harlow}},
  \bibinfo{journal}{Materials research bulletin} \textbf{\bibinfo{volume}{29}},
  \bibinfo{pages}{645} (\bibinfo{year}{1994}).

\bibitem[{\citenamefont{Kim et~al.}(2008)\citenamefont{Kim, Jin, Moon, Kim,
  Park, Leem, Yu, Noh, Kim, Oh et~al.}}]{kim2008novel}
\bibinfo{author}{\bibfnamefont{B.}~\bibnamefont{Kim}},
  \bibinfo{author}{\bibfnamefont{H.}~\bibnamefont{Jin}},
  \bibinfo{author}{\bibfnamefont{S.}~\bibnamefont{Moon}},
  \bibinfo{author}{\bibfnamefont{J.-Y.} \bibnamefont{Kim}},
  \bibinfo{author}{\bibfnamefont{B.-G.} \bibnamefont{Park}},
  \bibinfo{author}{\bibfnamefont{C.}~\bibnamefont{Leem}},
  \bibinfo{author}{\bibfnamefont{J.}~\bibnamefont{Yu}},
  \bibinfo{author}{\bibfnamefont{T.}~\bibnamefont{Noh}},
  \bibinfo{author}{\bibfnamefont{C.}~\bibnamefont{Kim}},
  \bibinfo{author}{\bibfnamefont{S.-J.} \bibnamefont{Oh}},
  \bibnamefont{et~al.}, \bibinfo{journal}{Physical review letters}
  \textbf{\bibinfo{volume}{101}}, \bibinfo{pages}{076402}
  (\bibinfo{year}{2008}).

\bibitem[{\citenamefont{Moon et~al.}(2008)\citenamefont{Moon, Jin, Kim, Choi,
  Lee, Yu, Cao, Sumi, Funakubo, Bernhard et~al.}}]{moon2008dimensionality}
\bibinfo{author}{\bibfnamefont{S.}~\bibnamefont{Moon}},
  \bibinfo{author}{\bibfnamefont{H.}~\bibnamefont{Jin}},
  \bibinfo{author}{\bibfnamefont{K.~W.} \bibnamefont{Kim}},
  \bibinfo{author}{\bibfnamefont{W.}~\bibnamefont{Choi}},
  \bibinfo{author}{\bibfnamefont{Y.}~\bibnamefont{Lee}},
  \bibinfo{author}{\bibfnamefont{J.}~\bibnamefont{Yu}},
  \bibinfo{author}{\bibfnamefont{G.}~\bibnamefont{Cao}},
  \bibinfo{author}{\bibfnamefont{A.}~\bibnamefont{Sumi}},
  \bibinfo{author}{\bibfnamefont{H.}~\bibnamefont{Funakubo}},
  \bibinfo{author}{\bibfnamefont{C.}~\bibnamefont{Bernhard}},
  \bibnamefont{et~al.}, \bibinfo{journal}{Physical review letters}
  \textbf{\bibinfo{volume}{101}}, \bibinfo{pages}{226402}
  (\bibinfo{year}{2008}).

\bibitem[{\citenamefont{Okada et~al.}(2013)\citenamefont{Okada, Walkup, Lin,
  Dhital, Chang, Khadka, Zhou, Jeng, Paranjape, Bansil
  et~al.}}]{okada2013imaging}
\bibinfo{author}{\bibfnamefont{Y.}~\bibnamefont{Okada}},
  \bibinfo{author}{\bibfnamefont{D.}~\bibnamefont{Walkup}},
  \bibinfo{author}{\bibfnamefont{H.}~\bibnamefont{Lin}},
  \bibinfo{author}{\bibfnamefont{C.}~\bibnamefont{Dhital}},
  \bibinfo{author}{\bibfnamefont{T.-R.} \bibnamefont{Chang}},
  \bibinfo{author}{\bibfnamefont{S.}~\bibnamefont{Khadka}},
  \bibinfo{author}{\bibfnamefont{W.}~\bibnamefont{Zhou}},
  \bibinfo{author}{\bibfnamefont{H.-T.} \bibnamefont{Jeng}},
  \bibinfo{author}{\bibfnamefont{M.}~\bibnamefont{Paranjape}},
  \bibinfo{author}{\bibfnamefont{A.}~\bibnamefont{Bansil}},
  \bibnamefont{et~al.}, \bibinfo{journal}{Nature materials}
  \textbf{\bibinfo{volume}{12}}, \bibinfo{pages}{707} (\bibinfo{year}{2013}).

\bibitem[{\citenamefont{Chen et~al.}(2015)\citenamefont{Chen, Hogan, Walkup,
  Zhou, Pokharel, Yao, Tian, Ward, Zhao, Parshall et~al.}}]{chen2015influence}
\bibinfo{author}{\bibfnamefont{X.}~\bibnamefont{Chen}},
  \bibinfo{author}{\bibfnamefont{T.}~\bibnamefont{Hogan}},
  \bibinfo{author}{\bibfnamefont{D.}~\bibnamefont{Walkup}},
  \bibinfo{author}{\bibfnamefont{W.}~\bibnamefont{Zhou}},
  \bibinfo{author}{\bibfnamefont{M.}~\bibnamefont{Pokharel}},
  \bibinfo{author}{\bibfnamefont{M.}~\bibnamefont{Yao}},
  \bibinfo{author}{\bibfnamefont{W.}~\bibnamefont{Tian}},
  \bibinfo{author}{\bibfnamefont{T.~Z.} \bibnamefont{Ward}},
  \bibinfo{author}{\bibfnamefont{Y.}~\bibnamefont{Zhao}},
  \bibinfo{author}{\bibfnamefont{D.}~\bibnamefont{Parshall}},
  \bibnamefont{et~al.}, \bibinfo{journal}{Physical Review B}
  \textbf{\bibinfo{volume}{92}}, \bibinfo{pages}{075125}
  (\bibinfo{year}{2015}).

\bibitem[{\citenamefont{Clancy et~al.}(2014)\citenamefont{Clancy, Lupascu,
  Gretarsson, Islam, Hu, Casa, Nelson, LaMarra, Cao, and
  Kim}}]{clancy2014dilute}
\bibinfo{author}{\bibfnamefont{J.}~\bibnamefont{Clancy}},
  \bibinfo{author}{\bibfnamefont{A.}~\bibnamefont{Lupascu}},
  \bibinfo{author}{\bibfnamefont{H.}~\bibnamefont{Gretarsson}},
  \bibinfo{author}{\bibfnamefont{Z.}~\bibnamefont{Islam}},
  \bibinfo{author}{\bibfnamefont{Y.}~\bibnamefont{Hu}},
  \bibinfo{author}{\bibfnamefont{D.}~\bibnamefont{Casa}},
  \bibinfo{author}{\bibfnamefont{C.}~\bibnamefont{Nelson}},
  \bibinfo{author}{\bibfnamefont{S.}~\bibnamefont{LaMarra}},
  \bibinfo{author}{\bibfnamefont{G.}~\bibnamefont{Cao}}, \bibnamefont{and}
  \bibinfo{author}{\bibfnamefont{Y.-J.} \bibnamefont{Kim}},
  \bibinfo{journal}{Physical Review B} \textbf{\bibinfo{volume}{89}},
  \bibinfo{pages}{054409} (\bibinfo{year}{2014}).

\bibitem[{\citenamefont{Calder et~al.}(2015)\citenamefont{Calder, Kim, Cao,
  Cantoni, May, Cao, Aczel, Matsuda, Choi, Haskel
  et~al.}}]{calder2015evolution}
\bibinfo{author}{\bibfnamefont{S.}~\bibnamefont{Calder}},
  \bibinfo{author}{\bibfnamefont{J.-W.} \bibnamefont{Kim}},
  \bibinfo{author}{\bibfnamefont{G.-X.} \bibnamefont{Cao}},
  \bibinfo{author}{\bibfnamefont{C.}~\bibnamefont{Cantoni}},
  \bibinfo{author}{\bibfnamefont{A.~F.} \bibnamefont{May}},
  \bibinfo{author}{\bibfnamefont{H.~B.} \bibnamefont{Cao}},
  \bibinfo{author}{\bibfnamefont{A.~A.} \bibnamefont{Aczel}},
  \bibinfo{author}{\bibfnamefont{M.}~\bibnamefont{Matsuda}},
  \bibinfo{author}{\bibfnamefont{Y.}~\bibnamefont{Choi}},
  \bibinfo{author}{\bibfnamefont{D.}~\bibnamefont{Haskel}},
  \bibnamefont{et~al.}, \bibinfo{journal}{Physical Review B}
  \textbf{\bibinfo{volume}{92}}, \bibinfo{pages}{165128}
  (\bibinfo{year}{2015}).

\bibitem[{\citenamefont{Kim et~al.}(2016)\citenamefont{Kim, Sung, Denlinger,
  and Kim}}]{kim2016observation}
\bibinfo{author}{\bibfnamefont{Y.}~\bibnamefont{Kim}},
  \bibinfo{author}{\bibfnamefont{N.}~\bibnamefont{Sung}},
  \bibinfo{author}{\bibfnamefont{J.}~\bibnamefont{Denlinger}},
  \bibnamefont{and} \bibinfo{author}{\bibfnamefont{B.}~\bibnamefont{Kim}},
  \bibinfo{journal}{Nature Physics} \textbf{\bibinfo{volume}{12}},
  \bibinfo{pages}{37} (\bibinfo{year}{2016}).

\bibitem[{\citenamefont{Qi et~al.}(2012)\citenamefont{Qi, Korneta, Li,
  Butrouna, Cao, Wan, Schlottmann, Kaul, and Cao}}]{qi2012spin}
\bibinfo{author}{\bibfnamefont{T.}~\bibnamefont{Qi}},
  \bibinfo{author}{\bibfnamefont{O.}~\bibnamefont{Korneta}},
  \bibinfo{author}{\bibfnamefont{L.}~\bibnamefont{Li}},
  \bibinfo{author}{\bibfnamefont{K.}~\bibnamefont{Butrouna}},
  \bibinfo{author}{\bibfnamefont{V.}~\bibnamefont{Cao}},
  \bibinfo{author}{\bibfnamefont{X.}~\bibnamefont{Wan}},
  \bibinfo{author}{\bibfnamefont{P.}~\bibnamefont{Schlottmann}},
  \bibinfo{author}{\bibfnamefont{R.}~\bibnamefont{Kaul}}, \bibnamefont{and}
  \bibinfo{author}{\bibfnamefont{G.}~\bibnamefont{Cao}},
  \bibinfo{journal}{Physical Review B} \textbf{\bibinfo{volume}{86}},
  \bibinfo{pages}{125105} (\bibinfo{year}{2012}).

\bibitem[{\citenamefont{De~La~Torre et~al.}(2015)\citenamefont{De~La~Torre,
  Walker, Bruno, Ricc{\'o}, Wang, Lezama, Scheerer, Giriat, Jaccard, Berthod
  et~al.}}]{de2015collapse}
\bibinfo{author}{\bibfnamefont{A.}~\bibnamefont{De~La~Torre}},
  \bibinfo{author}{\bibfnamefont{S.~M.} \bibnamefont{Walker}},
  \bibinfo{author}{\bibfnamefont{F.~Y.} \bibnamefont{Bruno}},
  \bibinfo{author}{\bibfnamefont{S.}~\bibnamefont{Ricc{\'o}}},
  \bibinfo{author}{\bibfnamefont{Z.}~\bibnamefont{Wang}},
  \bibinfo{author}{\bibfnamefont{I.~G.} \bibnamefont{Lezama}},
  \bibinfo{author}{\bibfnamefont{G.}~\bibnamefont{Scheerer}},
  \bibinfo{author}{\bibfnamefont{G.}~\bibnamefont{Giriat}},
  \bibinfo{author}{\bibfnamefont{D.}~\bibnamefont{Jaccard}},
  \bibinfo{author}{\bibfnamefont{C.}~\bibnamefont{Berthod}},
  \bibnamefont{et~al.}, \bibinfo{journal}{Physical review letters}
  \textbf{\bibinfo{volume}{115}}, \bibinfo{pages}{176402}
  (\bibinfo{year}{2015}).

\bibitem[{\citenamefont{Hogan et~al.}(2015)\citenamefont{Hogan, Yamani, Walkup,
  Chen, Dally, Ward, Dean, Hill, Islam, Madhavan et~al.}}]{hogan2015first}
\bibinfo{author}{\bibfnamefont{T.}~\bibnamefont{Hogan}},
  \bibinfo{author}{\bibfnamefont{Z.}~\bibnamefont{Yamani}},
  \bibinfo{author}{\bibfnamefont{D.}~\bibnamefont{Walkup}},
  \bibinfo{author}{\bibfnamefont{X.}~\bibnamefont{Chen}},
  \bibinfo{author}{\bibfnamefont{R.}~\bibnamefont{Dally}},
  \bibinfo{author}{\bibfnamefont{T.~Z.} \bibnamefont{Ward}},
  \bibinfo{author}{\bibfnamefont{M.}~\bibnamefont{Dean}},
  \bibinfo{author}{\bibfnamefont{J.}~\bibnamefont{Hill}},
  \bibinfo{author}{\bibfnamefont{Z.}~\bibnamefont{Islam}},
  \bibinfo{author}{\bibfnamefont{V.}~\bibnamefont{Madhavan}},
  \bibnamefont{et~al.}, \bibinfo{journal}{Physical review letters}
  \textbf{\bibinfo{volume}{114}}, \bibinfo{pages}{257203}
  (\bibinfo{year}{2015}).

\bibitem[{\citenamefont{Dhital et~al.}(2014)\citenamefont{Dhital, Hogan, Zhou,
  Chen, Ren, Pokharel, Okada, Heine, Tian, Yamani et~al.}}]{dhital2014carrier}
\bibinfo{author}{\bibfnamefont{C.}~\bibnamefont{Dhital}},
  \bibinfo{author}{\bibfnamefont{T.}~\bibnamefont{Hogan}},
  \bibinfo{author}{\bibfnamefont{W.}~\bibnamefont{Zhou}},
  \bibinfo{author}{\bibfnamefont{X.}~\bibnamefont{Chen}},
  \bibinfo{author}{\bibfnamefont{Z.}~\bibnamefont{Ren}},
  \bibinfo{author}{\bibfnamefont{M.}~\bibnamefont{Pokharel}},
  \bibinfo{author}{\bibfnamefont{Y.}~\bibnamefont{Okada}},
  \bibinfo{author}{\bibfnamefont{M.}~\bibnamefont{Heine}},
  \bibinfo{author}{\bibfnamefont{W.}~\bibnamefont{Tian}},
  \bibinfo{author}{\bibfnamefont{Z.}~\bibnamefont{Yamani}},
  \bibnamefont{et~al.}, \bibinfo{journal}{Nature communications}
  \textbf{\bibinfo{volume}{5}} (\bibinfo{year}{2014}).

\bibitem[{\citenamefont{Li et~al.}(2013)\citenamefont{Li, Kong, Qi, Jin, Yuan,
  DeLong, Schlottmann, and Cao}}]{li2013tuning}
\bibinfo{author}{\bibfnamefont{L.}~\bibnamefont{Li}},
  \bibinfo{author}{\bibfnamefont{P.}~\bibnamefont{Kong}},
  \bibinfo{author}{\bibfnamefont{T.}~\bibnamefont{Qi}},
  \bibinfo{author}{\bibfnamefont{C.}~\bibnamefont{Jin}},
  \bibinfo{author}{\bibfnamefont{S.}~\bibnamefont{Yuan}},
  \bibinfo{author}{\bibfnamefont{L.}~\bibnamefont{DeLong}},
  \bibinfo{author}{\bibfnamefont{P.}~\bibnamefont{Schlottmann}},
  \bibnamefont{and} \bibinfo{author}{\bibfnamefont{G.}~\bibnamefont{Cao}},
  \bibinfo{journal}{Physical Review B} \textbf{\bibinfo{volume}{87}},
  \bibinfo{pages}{235127} (\bibinfo{year}{2013}).

\bibitem[{\citenamefont{He et~al.}(2015)\citenamefont{He, Hogan, Mion, Hafiz,
  He, Denlinger, Mo, Dhital, Chen, Lin et~al.}}]{he2015spectroscopic}
\bibinfo{author}{\bibfnamefont{J.}~\bibnamefont{He}},
  \bibinfo{author}{\bibfnamefont{T.}~\bibnamefont{Hogan}},
  \bibinfo{author}{\bibfnamefont{T.~R.} \bibnamefont{Mion}},
  \bibinfo{author}{\bibfnamefont{H.}~\bibnamefont{Hafiz}},
  \bibinfo{author}{\bibfnamefont{Y.}~\bibnamefont{He}},
  \bibinfo{author}{\bibfnamefont{J.}~\bibnamefont{Denlinger}},
  \bibinfo{author}{\bibfnamefont{S.}~\bibnamefont{Mo}},
  \bibinfo{author}{\bibfnamefont{C.}~\bibnamefont{Dhital}},
  \bibinfo{author}{\bibfnamefont{X.}~\bibnamefont{Chen}},
  \bibinfo{author}{\bibfnamefont{Q.}~\bibnamefont{Lin}}, \bibnamefont{et~al.},
  \bibinfo{journal}{Nature materials} \textbf{\bibinfo{volume}{14}},
  \bibinfo{pages}{577} (\bibinfo{year}{2015}).

\bibitem[{\citenamefont{Wang and Senthil}(2011)}]{wang2011twisted}
\bibinfo{author}{\bibfnamefont{F.}~\bibnamefont{Wang}} \bibnamefont{and}
  \bibinfo{author}{\bibfnamefont{T.}~\bibnamefont{Senthil}},
  \bibinfo{journal}{Physical Review Letters} \textbf{\bibinfo{volume}{106}},
  \bibinfo{pages}{136402} (\bibinfo{year}{2011}).

\bibitem[{\citenamefont{Ahn et~al.}(2016)\citenamefont{Ahn, Song, Hogan,
  Wilson, and Moon}}]{ahn2016infrared}
\bibinfo{author}{\bibfnamefont{G.}~\bibnamefont{Ahn}},
  \bibinfo{author}{\bibfnamefont{S.}~\bibnamefont{Song}},
  \bibinfo{author}{\bibfnamefont{T.}~\bibnamefont{Hogan}},
  \bibinfo{author}{\bibfnamefont{S.}~\bibnamefont{Wilson}}, \bibnamefont{and}
  \bibinfo{author}{\bibfnamefont{S.}~\bibnamefont{Moon}},
  \bibinfo{journal}{Scientific Reports} \textbf{\bibinfo{volume}{6}},
  \bibinfo{pages}{32632} (\bibinfo{year}{2016}).

\bibitem[{\citenamefont{Hogan et~al.}(2016{\natexlab{a}})\citenamefont{Hogan,
  Dally, Upton, Clancy, Finkelstein, Kim, Graf, and
  Wilson}}]{hogan2016disordered}
\bibinfo{author}{\bibfnamefont{T.}~\bibnamefont{Hogan}},
  \bibinfo{author}{\bibfnamefont{R.}~\bibnamefont{Dally}},
  \bibinfo{author}{\bibfnamefont{M.}~\bibnamefont{Upton}},
  \bibinfo{author}{\bibfnamefont{J.}~\bibnamefont{Clancy}},
  \bibinfo{author}{\bibfnamefont{K.}~\bibnamefont{Finkelstein}},
  \bibinfo{author}{\bibfnamefont{Y.-J.} \bibnamefont{Kim}},
  \bibinfo{author}{\bibfnamefont{M.~J.} \bibnamefont{Graf}}, \bibnamefont{and}
  \bibinfo{author}{\bibfnamefont{S.~D.} \bibnamefont{Wilson}},
  \bibinfo{journal}{Physical Review B} \textbf{\bibinfo{volume}{94}},
  \bibinfo{pages}{100401} (\bibinfo{year}{2016}{\natexlab{a}}).

\bibitem[{\citenamefont{Sala et~al.}(2015)\citenamefont{Sala, Schnells,
  Boseggia, Simonelli, Al-Zein, Vale, Paolasini, Hunter, Perry, Prabhakaran
  et~al.}}]{sala2015evidence}
\bibinfo{author}{\bibfnamefont{M.~M.} \bibnamefont{Sala}},
  \bibinfo{author}{\bibfnamefont{V.}~\bibnamefont{Schnells}},
  \bibinfo{author}{\bibfnamefont{S.}~\bibnamefont{Boseggia}},
  \bibinfo{author}{\bibfnamefont{L.}~\bibnamefont{Simonelli}},
  \bibinfo{author}{\bibfnamefont{A.}~\bibnamefont{Al-Zein}},
  \bibinfo{author}{\bibfnamefont{J.}~\bibnamefont{Vale}},
  \bibinfo{author}{\bibfnamefont{L.}~\bibnamefont{Paolasini}},
  \bibinfo{author}{\bibfnamefont{E.}~\bibnamefont{Hunter}},
  \bibinfo{author}{\bibfnamefont{R.}~\bibnamefont{Perry}},
  \bibinfo{author}{\bibfnamefont{D.}~\bibnamefont{Prabhakaran}},
  \bibnamefont{et~al.}, \bibinfo{journal}{Physical Review B}
  \textbf{\bibinfo{volume}{92}}, \bibinfo{pages}{024405}
  (\bibinfo{year}{2015}).

\bibitem[{\citenamefont{Ding et~al.}(2016)\citenamefont{Ding, Yang, Chen, Kim,
  Han, Luo, Feng, Upton, Casa, Kim et~al.}}]{ding2016pressure}
\bibinfo{author}{\bibfnamefont{Y.}~\bibnamefont{Ding}},
  \bibinfo{author}{\bibfnamefont{L.}~\bibnamefont{Yang}},
  \bibinfo{author}{\bibfnamefont{C.-C.} \bibnamefont{Chen}},
  \bibinfo{author}{\bibfnamefont{H.-S.} \bibnamefont{Kim}},
  \bibinfo{author}{\bibfnamefont{M.~J.} \bibnamefont{Han}},
  \bibinfo{author}{\bibfnamefont{W.}~\bibnamefont{Luo}},
  \bibinfo{author}{\bibfnamefont{Z.}~\bibnamefont{Feng}},
  \bibinfo{author}{\bibfnamefont{M.}~\bibnamefont{Upton}},
  \bibinfo{author}{\bibfnamefont{D.}~\bibnamefont{Casa}},
  \bibinfo{author}{\bibfnamefont{J.}~\bibnamefont{Kim}}, \bibnamefont{et~al.},
  \bibinfo{journal}{Physical review letters} \textbf{\bibinfo{volume}{116}},
  \bibinfo{pages}{216402} (\bibinfo{year}{2016}).

\bibitem[{\citenamefont{Lu et~al.}(2017)\citenamefont{Lu, McNally, Sala,
  Terzic, Upton, Casa, Ingold, Cao, and Schmitt}}]{lu2017doping}
\bibinfo{author}{\bibfnamefont{X.}~\bibnamefont{Lu}},
  \bibinfo{author}{\bibfnamefont{D.}~\bibnamefont{McNally}},
  \bibinfo{author}{\bibfnamefont{M.~M.} \bibnamefont{Sala}},
  \bibinfo{author}{\bibfnamefont{J.}~\bibnamefont{Terzic}},
  \bibinfo{author}{\bibfnamefont{M.}~\bibnamefont{Upton}},
  \bibinfo{author}{\bibfnamefont{D.}~\bibnamefont{Casa}},
  \bibinfo{author}{\bibfnamefont{G.}~\bibnamefont{Ingold}},
  \bibinfo{author}{\bibfnamefont{G.}~\bibnamefont{Cao}}, \bibnamefont{and}
  \bibinfo{author}{\bibfnamefont{T.}~\bibnamefont{Schmitt}},
  \bibinfo{journal}{Physical Review Letters} \textbf{\bibinfo{volume}{118}},
  \bibinfo{pages}{027202} (\bibinfo{year}{2017}).

\bibitem[{\citenamefont{Chu et~al.}(2017)\citenamefont{Chu, Zhao, de~la Torre,
  Hogan, Wilson, and Hsieh}}]{chu2017charge}
\bibinfo{author}{\bibfnamefont{H.}~\bibnamefont{Chu}},
  \bibinfo{author}{\bibfnamefont{L.}~\bibnamefont{Zhao}},
  \bibinfo{author}{\bibfnamefont{A.}~\bibnamefont{de~la Torre}},
  \bibinfo{author}{\bibfnamefont{T.}~\bibnamefont{Hogan}},
  \bibinfo{author}{\bibfnamefont{S.~D.} \bibnamefont{Wilson}},
  \bibnamefont{and} \bibinfo{author}{\bibfnamefont{D.}~\bibnamefont{Hsieh}},
  \bibinfo{journal}{Nat Mater} \textbf{\bibinfo{volume}{16}},
  \bibinfo{pages}{200} (\bibinfo{year}{2017}).

\bibitem[{\citenamefont{Dhital et~al.}(2012)\citenamefont{Dhital, Khadka,
  Yamani, de~la Cruz, Hogan, Disseler, Pokharel, Lukas, Tian, Opeil
  et~al.}}]{dhital2012spin}
\bibinfo{author}{\bibfnamefont{C.}~\bibnamefont{Dhital}},
  \bibinfo{author}{\bibfnamefont{S.}~\bibnamefont{Khadka}},
  \bibinfo{author}{\bibfnamefont{Z.}~\bibnamefont{Yamani}},
  \bibinfo{author}{\bibfnamefont{C.}~\bibnamefont{de~la Cruz}},
  \bibinfo{author}{\bibfnamefont{T.}~\bibnamefont{Hogan}},
  \bibinfo{author}{\bibfnamefont{S.}~\bibnamefont{Disseler}},
  \bibinfo{author}{\bibfnamefont{M.}~\bibnamefont{Pokharel}},
  \bibinfo{author}{\bibfnamefont{K.}~\bibnamefont{Lukas}},
  \bibinfo{author}{\bibfnamefont{W.}~\bibnamefont{Tian}},
  \bibinfo{author}{\bibfnamefont{C.}~\bibnamefont{Opeil}},
  \bibnamefont{et~al.}, \bibinfo{journal}{Physical Review B}
  \textbf{\bibinfo{volume}{86}}, \bibinfo{pages}{100401}
  (\bibinfo{year}{2012}).

\bibitem[{\citenamefont{Hogan et~al.}(2016{\natexlab{b}})\citenamefont{Hogan,
  Bjaalie, Zhao, Belvin, Wang, Van~de Walle, Hsieh, and
  Wilson}}]{hogan2016structural}
\bibinfo{author}{\bibfnamefont{T.}~\bibnamefont{Hogan}},
  \bibinfo{author}{\bibfnamefont{L.}~\bibnamefont{Bjaalie}},
  \bibinfo{author}{\bibfnamefont{L.}~\bibnamefont{Zhao}},
  \bibinfo{author}{\bibfnamefont{C.}~\bibnamefont{Belvin}},
  \bibinfo{author}{\bibfnamefont{X.}~\bibnamefont{Wang}},
  \bibinfo{author}{\bibfnamefont{C.~G.} \bibnamefont{Van~de Walle}},
  \bibinfo{author}{\bibfnamefont{D.}~\bibnamefont{Hsieh}}, \bibnamefont{and}
  \bibinfo{author}{\bibfnamefont{S.~D.} \bibnamefont{Wilson}},
  \bibinfo{journal}{Physical Review B} \textbf{\bibinfo{volume}{93}},
  \bibinfo{pages}{134110} (\bibinfo{year}{2016}{\natexlab{b}}).

\bibitem[{Sup()}]{SupplementalMaterials}
\bibinfo{note}{See Supplemental Materials for additional details}.

\bibitem[{\citenamefont{Zikovsky et~al.}(2011)\citenamefont{Zikovsky, Peterson,
  Wang, Frost, and Hoffmann}}]{zikovsky2011crystalplan}
\bibinfo{author}{\bibfnamefont{J.}~\bibnamefont{Zikovsky}},
  \bibinfo{author}{\bibfnamefont{P.~F.} \bibnamefont{Peterson}},
  \bibinfo{author}{\bibfnamefont{X.~P.} \bibnamefont{Wang}},
  \bibinfo{author}{\bibfnamefont{M.}~\bibnamefont{Frost}}, \bibnamefont{and}
  \bibinfo{author}{\bibfnamefont{C.}~\bibnamefont{Hoffmann}},
  \bibinfo{journal}{Journal of Applied Crystallography}
  \textbf{\bibinfo{volume}{44}}, \bibinfo{pages}{418} (\bibinfo{year}{2011}).

\bibitem[{\citenamefont{Schultz et~al.}(2014)\citenamefont{Schultz,
  J{\o}rgensen, Wang, Mikkelson, Mikkelson, Lynch, Peterson, Green, and
  Hoffmann}}]{schultz2014integration}
\bibinfo{author}{\bibfnamefont{A.~J.} \bibnamefont{Schultz}},
  \bibinfo{author}{\bibfnamefont{M.~R.~V.} \bibnamefont{J{\o}rgensen}},
  \bibinfo{author}{\bibfnamefont{X.}~\bibnamefont{Wang}},
  \bibinfo{author}{\bibfnamefont{R.~L.} \bibnamefont{Mikkelson}},
  \bibinfo{author}{\bibfnamefont{D.~J.} \bibnamefont{Mikkelson}},
  \bibinfo{author}{\bibfnamefont{V.~E.} \bibnamefont{Lynch}},
  \bibinfo{author}{\bibfnamefont{P.~F.} \bibnamefont{Peterson}},
  \bibinfo{author}{\bibfnamefont{M.~L.} \bibnamefont{Green}}, \bibnamefont{and}
  \bibinfo{author}{\bibfnamefont{C.~M.} \bibnamefont{Hoffmann}},
  \bibinfo{journal}{Journal of Applied Crystallography}
  \textbf{\bibinfo{volume}{47}}, \bibinfo{pages}{915} (\bibinfo{year}{2014}).

\bibitem[{\citenamefont{Schultz et~al.}(1984)\citenamefont{Schultz, Srinivasan,
  Teller, Williams, and Lukehart}}]{doi:10.1021/ja00316a031}
\bibinfo{author}{\bibfnamefont{A.~J.} \bibnamefont{Schultz}},
  \bibinfo{author}{\bibfnamefont{K.}~\bibnamefont{Srinivasan}},
  \bibinfo{author}{\bibfnamefont{R.~G.} \bibnamefont{Teller}},
  \bibinfo{author}{\bibfnamefont{J.~M.} \bibnamefont{Williams}},
  \bibnamefont{and} \bibinfo{author}{\bibfnamefont{C.~M.}
  \bibnamefont{Lukehart}}, \bibinfo{journal}{Journal of the American Chemical
  Society} \textbf{\bibinfo{volume}{106}}, \bibinfo{pages}{999}
  (\bibinfo{year}{1984}).

\bibitem[{\citenamefont{Sheldrick}(2008)}]{Sheldrick:sc5010}
\bibinfo{author}{\bibfnamefont{G.~M.} \bibnamefont{Sheldrick}},
  \bibinfo{journal}{Acta Crystallographica Section A}
  \textbf{\bibinfo{volume}{64}}, \bibinfo{pages}{112} (\bibinfo{year}{2008}).

\bibitem[{\citenamefont{Harter et~al.}(2015)\citenamefont{Harter, Niu, Woss,
  and Hsieh}}]{harter2015high}
\bibinfo{author}{\bibfnamefont{J.}~\bibnamefont{Harter}},
  \bibinfo{author}{\bibfnamefont{L.}~\bibnamefont{Niu}},
  \bibinfo{author}{\bibfnamefont{A.}~\bibnamefont{Woss}}, \bibnamefont{and}
  \bibinfo{author}{\bibfnamefont{D.}~\bibnamefont{Hsieh}},
  \bibinfo{journal}{Optics letters} \textbf{\bibinfo{volume}{40}},
  \bibinfo{pages}{4671} (\bibinfo{year}{2015}).

\bibitem[{\citenamefont{Cooper et~al.}(2002)\citenamefont{Cooper, Gould,
  Parsons, and Watkin}}]{cooper2002derivation}
\bibinfo{author}{\bibfnamefont{R.~I.} \bibnamefont{Cooper}},
  \bibinfo{author}{\bibfnamefont{R.~O.} \bibnamefont{Gould}},
  \bibinfo{author}{\bibfnamefont{S.}~\bibnamefont{Parsons}}, \bibnamefont{and}
  \bibinfo{author}{\bibfnamefont{D.~J.} \bibnamefont{Watkin}},
  \bibinfo{journal}{Journal of applied crystallography}
  \textbf{\bibinfo{volume}{35}}, \bibinfo{pages}{168} (\bibinfo{year}{2002}).

\bibitem[{\citenamefont{Hwang et~al.}(2013)\citenamefont{Hwang, Son, Zhang,
  Janotti, Van~de Walle, and Stemmer}}]{hwang2013structural}
\bibinfo{author}{\bibfnamefont{J.}~\bibnamefont{Hwang}},
  \bibinfo{author}{\bibfnamefont{J.}~\bibnamefont{Son}},
  \bibinfo{author}{\bibfnamefont{J.~Y.} \bibnamefont{Zhang}},
  \bibinfo{author}{\bibfnamefont{A.}~\bibnamefont{Janotti}},
  \bibinfo{author}{\bibfnamefont{C.~G.} \bibnamefont{Van~de Walle}},
  \bibnamefont{and} \bibinfo{author}{\bibfnamefont{S.}~\bibnamefont{Stemmer}},
  \bibinfo{journal}{Physical Review B} \textbf{\bibinfo{volume}{87}},
  \bibinfo{pages}{060101} (\bibinfo{year}{2013}).

\bibitem[{\citenamefont{Torchinsky et~al.}(2015)\citenamefont{Torchinsky, Chu,
  Zhao, Perkins, Sizyuk, Qi, Cao, and Hsieh}}]{torchinsky2015structural}
\bibinfo{author}{\bibfnamefont{D.}~\bibnamefont{Torchinsky}},
  \bibinfo{author}{\bibfnamefont{H.}~\bibnamefont{Chu}},
  \bibinfo{author}{\bibfnamefont{L.}~\bibnamefont{Zhao}},
  \bibinfo{author}{\bibfnamefont{N.}~\bibnamefont{Perkins}},
  \bibinfo{author}{\bibfnamefont{Y.}~\bibnamefont{Sizyuk}},
  \bibinfo{author}{\bibfnamefont{T.}~\bibnamefont{Qi}},
  \bibinfo{author}{\bibfnamefont{G.}~\bibnamefont{Cao}}, \bibnamefont{and}
  \bibinfo{author}{\bibfnamefont{D.}~\bibnamefont{Hsieh}},
  \bibinfo{journal}{Physical review letters} \textbf{\bibinfo{volume}{114}},
  \bibinfo{pages}{096404} (\bibinfo{year}{2015}).

\bibitem[{\citenamefont{Ye et~al.}(2013)\citenamefont{Ye, Chi, Chakoumakos,
  Fernandez-Baca, Qi, and Cao}}]{ye2013magnetic}
\bibinfo{author}{\bibfnamefont{F.}~\bibnamefont{Ye}},
  \bibinfo{author}{\bibfnamefont{S.}~\bibnamefont{Chi}},
  \bibinfo{author}{\bibfnamefont{B.~C.} \bibnamefont{Chakoumakos}},
  \bibinfo{author}{\bibfnamefont{J.~A.} \bibnamefont{Fernandez-Baca}},
  \bibinfo{author}{\bibfnamefont{T.}~\bibnamefont{Qi}}, \bibnamefont{and}
  \bibinfo{author}{\bibfnamefont{G.}~\bibnamefont{Cao}},
  \bibinfo{journal}{Physical Review B} \textbf{\bibinfo{volume}{87}},
  \bibinfo{pages}{140406} (\bibinfo{year}{2013}).

\bibitem[{\citenamefont{Dhital et~al.}(2013)\citenamefont{Dhital, Hogan,
  Yamani, de~la Cruz, Chen, Khadka, Ren, and Wilson}}]{dhital2013neutron}
\bibinfo{author}{\bibfnamefont{C.}~\bibnamefont{Dhital}},
  \bibinfo{author}{\bibfnamefont{T.}~\bibnamefont{Hogan}},
  \bibinfo{author}{\bibfnamefont{Z.}~\bibnamefont{Yamani}},
  \bibinfo{author}{\bibfnamefont{C.}~\bibnamefont{de~la Cruz}},
  \bibinfo{author}{\bibfnamefont{X.}~\bibnamefont{Chen}},
  \bibinfo{author}{\bibfnamefont{S.}~\bibnamefont{Khadka}},
  \bibinfo{author}{\bibfnamefont{Z.}~\bibnamefont{Ren}}, \bibnamefont{and}
  \bibinfo{author}{\bibfnamefont{S.~D.} \bibnamefont{Wilson}},
  \bibinfo{journal}{Physical Review B} \textbf{\bibinfo{volume}{87}},
  \bibinfo{pages}{144405} (\bibinfo{year}{2013}).

\bibitem[{\citenamefont{Momma and Izumi}(2011)}]{Momma2011}
\bibinfo{author}{\bibfnamefont{K.}~\bibnamefont{Momma}} \bibnamefont{and}
  \bibinfo{author}{\bibfnamefont{F.}~\bibnamefont{Izumi}},
  \bibinfo{journal}{Journal of Applied Crystallography}
  \textbf{\bibinfo{volume}{44}}, \bibinfo{pages}{1272} (\bibinfo{year}{2011}).

\end{thebibliography}
%
%
\end{document}